\documentclass[reprint,aps,prd,onecolumn,showpacs,nofootinbib,notitlepage,superscriptaddress]{revtex4-1}
\usepackage[T1]{fontenc}
\usepackage[latin9]{inputenc}
\usepackage{bm}
\usepackage{amsmath}
\usepackage{esint}
\usepackage{color}
\usepackage{graphicx}
\PassOptionsToPackage{normalem}{ulem}
\usepackage{ulem}

\usepackage{times}
\usepackage{hyperref,enumerate}

\begin{document}

\title{Derivative interactions for a spin-2 field at cubic order}

\author{Xian Gao}%
    \email[Email: ]{gao@th.phys.titech.ac.jp}
    \affiliation{%
        Department of Physics, Tokyo Institute of Technology,\\ 
        2-12-1 Ookayama, Meguro, Tokyo 152-8551, Japan}

\date{\today}

\begin{abstract}
Lorentz invariant derivative interactions for a single spin-2 field are investigated, up to the cubic order. 
We start from the most general Lorentz invariant terms involving two spacetime derivatives, which are polynomials in the spin-2 field as well as its first  derivatives.
Using a perturbative Arnowitt-Deser-Misner analysis, we determined the parameters such that the corresponding Hamiltonian possesses a Lagrange multiplier, which would signify there are at most 5 degrees of freedom that are propagating. 
The resulting derivative terms are linear combinations of terms coming from the expansion of the Einstein-Hilbert Lagrangian around a Minkowski background, as well as the cubic ``pseudolinear derivative term'' identified in Hinterbichler [J. High Energy Phys. 10 (\textbf{2013}) 102].
We also derived the compatible potential terms, which are linear combinations of the expansions of the first two de Rham-Gabadadze-Tolley mass terms in unitary gauge.
\end{abstract}

\maketitle

\section{Introduction}
\label{sec:intro}

Remarkable progress has been made in the study of ghost-free nonlinear massive gravity very recently.
In four dimensions, a massive spin-2 particle should carry 5 degrees of freedom and any consistent theory for massive gravity must satisfy this requirement. 
The quadratic order mass term was first constructed by Fierz and Pauli (FP) \cite{Fierz:1939ix}, which together with the linearized Einstein-Hilbert term describes a free massive spin-2 particle propagating on a Minkowski background.
In the FP theory, however, the helicity-0 mode couples to the trace of the matter energy-momentum tensor with the same strength as the helicity-2 modes, which prevents the theory from recovering linearized general relativity (GR) in the massless limit, known as the van Dam-Veltman-Zakharov discontinuity \cite{vanDam:1970vg,Zakharov:1970cc}.
It was argued that this discontinuity could be avoided through the Vainshtein mechanism \cite{Vainshtein:1972sx}, where the nonlinear interactions become important in the massless limit.
However, when accompanied with the fully nonlinear Einstein-Hilbert term, the Fierz-Pauli mass term as well as its ``naive'' higher order generalizations will propagate a sixth degree of freedom at nonlinear orders, known as the Boulware-Deser (BD) ghost \cite{Boulware:1973my,Boulware:1974sr}, which implies the classical instability of the theory.

Counting degrees of freedom can be systematically performed in the Hamiltonian analysis. 
For GR, the existence of the Hamiltonian and momentum constraints is transparent in the Arnowitt-Deser-Misner (ADM) formalism, where the lapse $N$ and shift $N_i$ enter the Hamiltonian linearly. For massive gravity, however, things become subtle. 
First, for a general mass term, which could be \textit{a priori} an arbitrary function of the metric, the lapse and shift appear nonlinearly and thus no longer act directly as Lagrange multipliers \cite{Creminelli:2005qk}.
On the other hand, there is no reason that the constraint must only be generated by the lapse itself. Instead, it may be generated by the set of lapse and shift together, i.e., some combination of the lapse and the shift. 
This possibility was realized only very recently. Using the nonlinear St\"{u}ckelberg method developed in \cite{ArkaniHamed:2002sp}, consistent generalizations of the FP term were constructed up to the quintic order in \cite{deRham:2010ik}, and resummed into a fully nonlinear form by de Rham, Gabadadze, and Tolley (dRGT) \cite{deRham:2010kj} (see \cite{deRham:2014zqa,Hinterbichler:2011tt,Rubakov:2008nh} for recent reviews). The dRGT theory possesses a Hamiltonian constraint (as well as a secondary constraint) necessary to remove the BD ghost \cite{Hassan:2011hr,Hassan:2011tf} (see also \cite{Deffayet:2012nr} for the constraint analysis in a covariant manner).

The dRGT mass terms are determined under the assumption that the kinetic term for gravity is GR, which is the unique theory for a single massless spin-2 field. 
However, for a massive spin-2 field, diffeomorphism invariance is broken and thus one is allowed to consider other diffeomorphism noninvariant kinetic (derivative) terms.
Conversely, if the kinetic term is different from GR, there is a possibility that the mass term compatible with this non-GR kinetic term is also different from the dRGT mass term. If this is true, we will have a more general class of massive gravity theories beyond the dRGT one.
Moreover, in bi/multimetric theories (e.g., \cite{Hassan:2011zd}), metrics interact with each other only through potential terms, which take the dRGT form. The non-GR derivative terms, if they exist, may also induce derivative interactions among different metrics, and thus extend our understanding of interacting spin-2 fields.
This work is devoted to exploring this possibility.

In \cite{Folkerts:2011ev} (see also \cite{Folkerts:2013mra}), non-GR cubic self-interaction terms with two spacetime derivatives was shown to exist using the helicity decomposition. Such a possibility was systematically investigated in \cite{Hinterbichler:2013eza} in arbitrary dimensions and with more than two  derivatives, where a class of so-called ``pseudolinear'' terms was identified.
The pseudolinear terms are nonlinear in $h_{\mu\nu}$, but are invariant under the linearized gauge symmetries. 
In $d$ dimensions, there are $(d+1)$ mass terms satisfying this property,
	\[
		\mathcal{L}_{0,n} \sim \delta^{\mu_1 \cdots \mu_n}_{\nu_1 \cdots \nu_n} h^{\nu_1}_{\mu_1} \cdots h^{\nu_n}_{\mu_n},\qquad n=0,1,\dots, d,
	\]
where the FP term is just $\mathcal{L}_{0,2}$, while $\mathcal{L}_{0,n}$ with $n\geq 3$ are the higher order analogue of FP term.
The pseudolinear terms with two derivatives are 
	\[
		\mathcal{L}_{2,n} \sim \delta^{\mu_1 \cdots \mu_{n+1}}_{\nu_1 \cdots \nu_{n+1}}  \partial_{\mu_1} \partial^{\nu_1} h^{\nu_2}_{\mu_2} \cdots h^{\nu_{n+1}}_{\mu_{n+1}}, \qquad n=1,\dots, d-1,
	\]
where the linearized Einstein-Hilbert term is nothing but $\mathcal{L}_{2,2}$, while the non-GR cubic term found in \cite{Folkerts:2011ev} just corresponds to $\mathcal{L}_{2,3}$ and is the unique pseudolinear derivative term in four dimensions.
In four dimensions, $\mathcal{L}_{0,2}$, $\mathcal{L}_{0,3}$ and $\mathcal{L}_{0,4}$ are the leading terms in the expansion of the corresponding dRGT mass terms around a Minkowski background.
Conversely, the full dRGT mass terms can be viewed as the ``nonlinear completion'' of these pseudolinear mass terms.
It was thus conjectured that there is also a nonlinear completion of the pseudolinear derivative term $\mathcal{L}_{2,3}$ \cite{Hinterbichler:2013eza}.

Two types of nonlinear derivative terms were constructed in \cite{Kimura:2013ika}, $G^{\mu\nu}\mathcal{K}_{\mu\nu}$ and ${}^{\ast} R^{\mu\nu\rho\sigma}\mathcal{K}_{\mu\rho}\mathcal{K}_{\nu\sigma}$ 
(with $\mathcal{K}_{\nu}^{\mu}\equiv\delta_{\nu}^{\mu}-\sqrt{\delta_{\nu}^{\mu}-H_{\nu}^{\mu}}$, where $H_{\mu\nu}$ is the covariantized metric perturbation $h_{\mu\nu}$ that reduces to $h_{\mu\nu}$ in unitary gauge; see, e.g., \cite{deRham:2010ik,deRham:2010kj} for details), which reproduce the linearized Einstein-Hilbert term $\mathcal{L}_{2,2}$ and the pseudolinear derivative term $\mathcal{L}_{2,3}$, respectively, when being expanded around the Minkowski background. 
However, both terms were shown to suffer from a ghost at the energy scale $\Lambda_3$ in the decoupling limit. 
The same two nonlinear derivative terms were derived from a higher dimensional Gauss-Bonnet term in \cite{deRham:2013tfa} using the ``dimensional deconstruction'' approach \cite{deRham:2013awa}, which has successfully reproduced the dRGT mass terms  from a higher dimensional Einstein-Hilbert term.\footnote{Alternative terms consisting of $\mathcal{K}_{\mu\nu}$ and its covariant derivatives are studied in \cite{Ohara:2014vua}, which also have the ghost problem.}
Based on a perturbative analysis, a ``no-go'' theorem was further proposed in \cite{deRham:2013tfa}, which claims that the pseudolinear derivative term $\mathcal{L}_{2,3}$ does not possess a ghost-free nonlinear completion, and thus there is no diffeomorphism-breaking but Lorentz invariant  derivative terms in metric formulation of massive gravity.

In this article, we perform an alternative perturbative ADM analysis to find possibly ghost-free derivative terms, up to the cubic order. 
We restrict ourselves to the case of two spacetime derivatives 
and consider the most general Lorentz invariant two-derivative terms, which are polynomials in the spin-2 field $h_{\mu\nu}$ as well as its first derivatives. They describe a single massive spin-2 particle propagating on a Minkowski background. 
By tuning the parameters such that the Hamiltonian possesses a possible constraint that eliminates the ghost degree of freedom, we can determine the derivative terms as well as the compatible potential terms.
In general, as is for the dRGT terms \cite{Hassan:2011hr,Hassan:2011tf}, such a constraint is generated by a combination of $\{\delta N,N_i\}$, which are the perturbation of the lapse function and the shift vector, respectively.
Our investigation can be viewed as a complementary but more systematic approach with respect to the analysis in \cite{deRham:2013tfa}, where first a nonlinear St\"{u}ckelberg analysis was performed to determine the necessary choice of parameters, whose sufficiency was subsequently tested by a specific perturbative ADM analysis.

In Sec.\ref{sec:pert_Ham}, we set up the necessary formalism for our perturbative ADM analysis for the Hamiltonian, and in Sec.\ref{sec:mass_term}, as an example, we show how the dRGT mass terms can be determined perturbatively in our approach. In Sec.\ref{sec:der}, we determine the derivative terms as well as the compatible mass terms up to the cubic order. We summarize our results in Sec.\ref{sec:con}.

\section{Perturbative Hamiltonian}\label{sec:pert_Ham}

In this work, we will deal with Lagrangian consisting of Lorentz invariant polynomials of a spin-2 field $h_{\mu\nu}$ and its first spacetime derivatives.
Moreover, we restrict ourselves to the case with no more than two spacetime derivatives.

The derivation of Hamiltonian relies on the splitting of space and time. Thus it is convenient to use the ADM variables $\{ N,N_{i},\gamma_{ij}\}$, which are defined through
	\begin{equation}\label{metric_ADM}
		ds^{2}=g_{\mu\nu}dx^{\mu}dx^{\nu}=-\left(N^{2}-\gamma^{ij}N_{i}N_{j}\right)dtdt+2N_{i}dtdx^{i}+\gamma_{ij}dx^{i}dx^{j},
	\end{equation}
where $N$ is the lapse, $N_i$ is the shift, and $\gamma^{ij}$ is the matrix inverse of the spatial metric $\gamma_{ij}$.
The deviation from a Minkowski background is parametrized by
	\begin{equation}
		N\equiv 1+\alpha,\qquad N_{i}\equiv\beta_{i},
	\end{equation}
and $\gamma_{ij}\equiv\delta_{ij}+h_{ij}$, so that
	\begin{equation}
		\gamma^{ij}=\delta_{ij}-h_{ij}+h_{ik}h_{kj}+\mathcal{O}\left(h^{3}\right),
	\end{equation}
where throughout this work repeated lower spatial indices are summed using $\delta_{ij}$.
We use $\{\alpha,\beta_i,h_{ij}\}$ as the perturbative variables, for example, in terms of which $h_{00}$ can be rewritten as
	\begin{equation}
		h_{00}= -\alpha\left(2+\alpha\right)+\left(\delta_{ij}-h_{ij}\right)\beta_{i}\beta_{j}+\cdots,
	\end{equation}
where ``$\cdots$'' denotes terms that are quartic (and higher) order in $\beta_i$ and $h_{ij}$.

No matter whether massless or massive, it is the spatial component $h_{ij}$ that carries the dynamical degrees of freedom in $h_{\mu\nu}\equiv \delta g_{\mu\nu}$. This means first we have to make sure that the Lagrangian can be expressed in the first order form, where only $h_{ij}$ has time derivatives, while $\alpha$ and $\beta_i$ have vanishing conjugate momenta and thus do not enter the phase space. As we will see, at least up to the cubic order, this simple requirement has already determined the consistent derivative terms.\footnote{While additional constraints, including the consistency conditions between the potential and derivative terms, must be included in order to determine the potential terms.} 

After eliminating time derivatives on $\alpha$ and $\beta_{i}$, we will deal with a Lagrangian taking the following form:
	\begin{equation}
		\mathcal{L}^{\text{ADM}}=\frac{1}{2}\mathcal{G}_{ij,kl}\dot{h}_{ij}\dot{h}_{kl}+\mathcal{F}_{ij}\dot{h}_{ij}+\mathcal{W},\label{L_ADM_stru}
	\end{equation}
where a dot denotes time derivative, and $\mathcal{G}$, $\mathcal{F}$, and $\mathcal{W}$ are functions of $\{\alpha,\beta_i,h_{ij}\}$ containing no time derivatives, schematically,
	\begin{equation}
		\mathcal{G}_{ij,kl}=\mathcal{G}_{ij,kl}\left(\alpha,\beta,h\right),\qquad
		\mathcal{F}_{ij} = \mathcal{F}_{ij}\left(\alpha,\beta,h\right),\qquad
		\mathcal{W} = \mathcal{W}\left(\alpha,\beta,h\right).\label{GFW_gen}
	\end{equation}
Note $\mathcal{F}_{ij}$ is symmetric with respect to its indices, while $\mathcal{G}_{ij,kl}$ has the following symmetries
	\begin{equation}
		\mathcal{G}_{ij,kl}=\mathcal{G}_{ji,kl}=\mathcal{G}_{ij,lk}=\mathcal{G}_{kl,ij}.\label{G_symm}
	\end{equation}
The inverse of $\mathcal{G}_{ij,kl}$ is defined as
	\begin{equation}
	\mathcal{G}_{ij,kl}^{-1}\mathcal{G}_{kl,i'j'}=\mathbf{I}_{ij,i'j'},\label{G_inverse_def}
	\end{equation}
where $\mathbf{I}_{ij,i'j'}$ is the identity in the space of symmetric matrices,
	\begin{equation}
	\mathbf{I}_{ij,i'j'}\equiv\frac{1}{2}\left(\delta_{ii'}\delta_{jj'}+\delta_{ij'}\delta_{ji'}\right),\label{identity_def}
	\end{equation}
which has the same symmetries as in (\ref{G_symm}).

The conjugate momentum of $h_{ij}$ is defined by
	\begin{equation}
	\pi_{ij}\equiv\frac{\partial\mathcal{L}^{\mathrm{ADM}}}{\partial\dot{h}_{ij}}=\mathcal{G}_{ij,kl}\dot{h}_{kl}+\mathcal{F}_{ij},\label{pi_gen}
	\end{equation}
from which we solve
	\begin{equation}
	\dot{h}_{ij}=\mathcal{G}_{ij,kl}^{-1}\left(\pi_{kl}-\mathcal{F}_{kl}\right).\label{hdot_pi_gen}
	\end{equation}
The Hamiltonian density is thus given by
	\begin{equation}
		\mathcal{H} \equiv \pi_{ij}\dot{h}_{ij}-\mathcal{L} = \frac{1}{2}\left(\pi_{ij}-\mathcal{F}_{ij}\right)\mathcal{G}_{ij,kl}^{-1}\left(\pi_{kl}-\mathcal{F}_{kl}\right)-\mathcal{W}.\label{Hamiltonian_gen}
	\end{equation}
In this work, we are dealing with perturbative theories, where $\mathcal{G}_{ij,kl}$, etc., are polynomials of $\{\alpha,\beta_i,h_{ij}\}$ and can be expanded as
	\begin{eqnarray}
	\mathcal{G}_{ij,kl} & = & \mathcal{G}_{ij,kl}^{(0)}+\mathcal{G}_{ij,kl}^{(1)}+\cdots,\label{G_pert_exp}\\
	\mathcal{F}_{ij} & = & \mathcal{F}_{ij}^{(1)}+\mathcal{F}_{ij}^{(2)}+\cdots,\label{F_pert_exp}\\
	\mathcal{W} & = & \mathcal{W}^{(2)}+\mathcal{W}^{(3)}+\cdots,\label{W_pert_exp}
	\end{eqnarray}
and
	\begin{equation}
		\mathcal{G}_{ij,kl}^{-1}=\left(\mathcal{G}^{-1}\right)_{ij,kl}^{(0)}+\left(\mathcal{G}^{-1}\right)_{ij,kl}^{(1)}+\cdots,
	\end{equation}
where $\left(\mathcal{G}^{-1}\right)_{ij,kl}^{(0)}\equiv\left(\mathcal{G}^{(0)}\right)_{ij,kl}^{-1}$, and superscript ``${}^{(n)}$'' denotes the order in perturbations.
The corresponding Hamiltonian density is also expanded as
	\begin{equation}
	\mathcal{H}=\mathcal{H}_{2}+\mathcal{H}_{3}+\mathcal{H}_{4}+\cdots,\label{Hamiltonaian_pert}
	\end{equation}
with, for example,
	\begin{eqnarray}
	\mathcal{H}_{2} & = & \frac{1}{2}\left(\pi_{ij}-\mathcal{F}_{ij}^{(1)}\right)\left(\mathcal{G}^{-1}\right)_{ij,kl}^{(0)}\left(\pi_{kl}-\mathcal{F}_{kl}^{(1)}\right)-\mathcal{W}^{(2)},\label{H2_gen}\\
	\mathcal{H}_{3} & = & \frac{1}{2}\left(\pi_{ij}-\mathcal{F}_{ij}^{(1)}\right)\left(\mathcal{G}^{-1}\right)_{ij,kl}^{(1)}\left(\pi_{kl}-\mathcal{F}_{kl}^{(1)}\right)-\mathcal{F}_{ij}^{(2)}\left(\mathcal{G}^{-1}\right)_{ij,kl}^{(0)}\left(\pi_{kl}-\mathcal{F}_{kl}^{(1)}\right)-\mathcal{W}^{(3)}.\label{H3_gen}
	\end{eqnarray}

For general functions $\mathcal{G}_{ij,k,l}$, etc., as in (\ref{GFW_gen}), neither $\alpha$ nor $\beta_i$ enter $\mathcal{H}$ linearly, or strictly speaking, the determinant of Hessian $\det \left( \frac{\partial^2\mathcal{H}}{\partial n_a \partial n_b} \right)$ with $n_a\equiv \{\alpha,\beta_i\}$ does not vanish. There is no further constraint, and thus the system will propagate all 6 degrees of freedom of $h_{ij}$, which signifies the existence of the BD ghost.
Fortunately and as we will see, it is possible to tune the parameters such that there exists a constraint generated by the set of four variables $\{\alpha,\beta_i\}$. This is also the essence
in proving the vanishing of the BD ghost for the dRGT mass terms \cite{Hassan:2011hr,Hassan:2011tf}.

\subsection{Linearized Einstein-Hilbert}

As is well known, the quadratic kinetic term is uniquely determined to be (see Appendix \ref{sec:linear_EH})
	\begin{equation}
	\mathcal{L}_{2}^{\mathrm{der}} = b_{1}\left(\partial_{\lambda}h^{\mu\nu}\partial^{\lambda}h_{\mu\nu}+2\,\partial_{\mu}h^{\mu\nu}\partial_{\nu}h-2\,\partial_{\nu}h^{\mu\nu}\partial^{\lambda}h_{\mu\lambda}-\,\partial_{\mu}h\partial^{\mu}h\right) \\
	  \simeq  -4b_{1}\mathcal{L}^{\mathrm{GR}}_2,\label{L2_EH}
	\end{equation}
where $\mathcal{L}^{\mathrm{GR}} = \sqrt{-g}R$ and a subscript ``${}_2$'' denotes the expansion at the quadratic order around the Minkowski background. Throughout this work, upper Lorentzian indices are raised by Minkowski metric $\eta^{\mu\nu}$.
In (\ref{L2_EH}) we keep the overall factor $b_1$ simply for later convenience.
In terms of perturbative ADM variables $\{\alpha,\beta_i,h_{ij}\}$, at the quadratic order, we have
	\begin{equation}
	\mathcal{L}_{2}^{\mathrm{der,ADM}}=\frac{1}{2}\mathcal{G}_{ij,kl}^{(0)}\dot{h}_{ij}\dot{h}_{kl}+\mathcal{F}_{ij}^{(1)}\dot{h}_{ij}+\mathcal{W}^{(2)},\label{L2_ADM_expl}
	\end{equation}
with
	\begin{eqnarray}
	\mathcal{G}_{ij,kl}^{(0)} & = & b_{1}\left(2\delta_{ij}\delta_{kl}-\delta_{ik}\delta_{jl}-\delta_{il}\delta_{jk}\right),\label{G0}\\
	\mathcal{F}_{ij}^{(1)} & = & 2b_{1}\left(\partial_{i}\beta_{j}+\partial_{j}\beta_{i}-2\partial_{k}\beta_{k}\delta_{ij}\right),\label{F1}\\
	\mathcal{W}^{(2)} & = & b_{1}\Big[-2\partial_{j}h_{ij}\partial_{k}h_{ik}+\partial_{j}h_{ii}\left(-\partial_{j}h_{kk}+2\partial_{k}h_{kj}\right)+\partial_{k}h_{ij}\partial_{k}h_{ij}\nonumber \\
	 &  & +2\left(\partial_{i}\beta_{i}\partial_{j}\beta_{j}-\partial_{j}\beta_{i}\partial_{j}\beta_{i}\right)+4\partial_{j}\alpha\left(\partial_{i}h_{ij}-\partial_{j}h_{ii}\right)\Big].\label{W2}
	\end{eqnarray}
At this point, we can easily determine
	\begin{equation}
		\left(\mathcal{G}^{(0)}\right)_{ij,kl}^{-1}= \frac{1}{4b_{1}}\left(\delta_{ij}\delta_{kl}-\delta_{ik}\delta_{jl}-\delta_{il}\delta_{jk}\right).\label{G0_inv}
	\end{equation}

Plugging the above into (\ref{H2_gen}) and after some manipulations, the corresponding Hamiltonian density is given by
	\begin{eqnarray}
	\mathcal{H}_{2}^{\mathrm{der}} & = & \frac{1}{8b_{1}}\left(\pi_{ii}\pi_{jj}-2\pi_{ij}\pi_{ij}\right)+2\pi_{ij}\partial_{i}\beta_{j}+4b_{1}\alpha\left(\partial_{i}\partial_{j}h_{ij}-\partial^{2}h_{ii}\right)\nonumber \\
	 &  & -b_{1}\left[-2\partial_{j}h_{ij}\partial_{k}h_{ik}+\partial_{j}h_{ii}\left(-\partial_{j}h_{kk}+2\partial_{k}h_{kj}\right)+\partial_{k}h_{ij}\partial_{k}h_{ij}\right],\label{H2_fin}
	\end{eqnarray}
up to total derivatives. Equation (\ref{H2_fin}) explicitly shows that, at the quadratic order, both $\alpha$ and $\beta_i$ enter the Hamiltonian linearly and thus act as Lagrange multipliers.

\subsection{Fierz-Pauli term and its cubic generalization}\label{sec:mass_term}

As a simple illustration of our strategy, now we show how the Fierz-Pauli term \cite{Fierz:1939ix} and its cubic generalization are determined by imposing the appropriate constraint structure.

The most general Lorentz invariant quadratic and cubic potential terms
are
	\begin{eqnarray}
		\mathcal{L}_{2}^{\mathrm{pot}} & = & b'_{1}h_{\mu}^{\mu}h_{\nu}^{\nu}+ b'_{2}h_{\mu\nu}h^{\mu\nu},\label{L2_mass}\\
		\mathcal{L}_{3}^{\mathrm{pot}} & = & c'_{1}h_{\mu}^{\mu}h_{\nu}^{\nu}h_{\rho}^{\rho}+c'_{2}h_{\rho}^{\rho}h_{\nu}^{\mu}h_{\mu}^{\nu}+c'_{3}h_{\nu}^{\mu}h_{\rho}^{\nu}h_{\mu}^{\rho},\label{L3_mass}
	\end{eqnarray}
where $b'_{i}$'s and $c'_{i}$'s are constant parameters to be determined. In
terms of ADM variables, at the quadratic order,
\begin{equation}
\mathcal{L}_{2}^{\mathrm{pot,ADM}}=4\left(b'_{1}+b'_{2}\right)\alpha^{2}+4b'_{1}\alpha h_{ii}-2b'_{2}\beta_{i}\beta_{i}+b'_{1}h_{ii}h_{jj}+b'_{2}h_{ij}h_{ij},\label{L2_mass_ADM}
\end{equation}
while at the cubic order,
	\begin{eqnarray}
		\mathcal{L}_{3}^{\mathrm{pot,ADM}} & = & 8\left(c'_{1}+c'_{2}+c'_{3}\right)\alpha^{3}+2\left(b'_{1}+6c'_{1}+2c'_{2}\right)\alpha^{2}h_{ii}-2\left(2c'_{2}+3c'_{3}\right)\alpha\beta_{i}\beta_{i}\nonumber\\
		 &  & +2\alpha\left(3c'_{1}h_{ii}h_{jj}+c'_{2}h_{ij}h_{ij}\right)-3c'_{3}h_{ij}\beta_{i}\beta_{j}-2\left(b'_{1}+c'_{2}\right)h_{ii}\beta_{j}\beta_{j}\nonumber\\
		 &  & +c'_{1}h_{ii}h_{jj}h_{kk}+c'_{2}h_{ii}h_{jk}h_{jk}+c'_{3}h_{ij}h_{jk}h_{ki}.\label{L3_mass_ADM}
	\end{eqnarray}
Note $\mathcal{L}_{2}^{\mathrm{pot}}$ also contributes to $\mathcal{L}_{3}^{\mathrm{pot,ADM}}$.

To guarantee the possible existence of a constraint relevant to $\alpha$, first the self-coupling terms of $\alpha$ must be vanishing, which implies, at the quadratic order,
	\begin{equation}
		b'_{1}+b'_{2} = 0,\label{b_mass_cons}
	\end{equation}
and, at the cubic order, 
	\begin{equation}
		c'_{1}+c'_{2}+c'_{3}=0.\label{c_mass_cons1}
	\end{equation}
At the quadratic order, (\ref{b_mass_cons}) uniquely fixes the mass terms to the form of Fierz-Pauli, up to an overall factor $b_1'$.
While at the cubic order, there are two apparently problematic terms in (\ref{L3_mass_ADM}), proportional to $\alpha^{2}h_{ii}$ and $\alpha\beta_{i}\beta_{i}$, respectively, which seem to prevent $\alpha$ from being a Lagrange multiplier.
These two terms, however, can be regrouped (together with the relevant terms in $\mathcal{L}_{2}^{\mathrm{pot,ADM}}$) into
	\begin{equation}
		2b'_1\left[\beta_{i}\beta_{i}-2\frac{2c'_{2}+3c'_{3}}{2b'_1}\alpha\beta_{i}\beta_{i}\right]\subset 2b'_1\left(\beta_{i}-\frac{2c'_{2}+3c'_{3}}{2b'_1}\alpha\beta_{i}\right)^{2} \label{beta_comb}
	\end{equation}
and
	\begin{equation}
		4b'_{1}\alpha h_{ii}+2\left(b'_{1}+6c'_{1}+2c'_{2}\right)\alpha^{2}h_{ii}=4b'_{1}\left(\alpha+\left(\frac{1}{2}+\frac{3c_{1}'+c_{2}'}{b_{1}'}\right)\alpha^{2}\right)h_{ii}.\label{alpha_comb}
	\end{equation}
Thus we can introduce new variables
	\begin{eqnarray}
		\tilde{\beta}_{i} & = & \beta_{i}-\frac{2c'_{2}+3c'_{3}}{2b'_{1}}\alpha\beta_{i},\label{beta_mass_redef}\\
		\tilde{\alpha} & = & \alpha+\left(\frac{1}{2}+\frac{3c'_{1}+c'_{2}}{b'_{1}}\right)\alpha^{2} .\label{alpha_mass_redef}
	\end{eqnarray}
With these redefined variables, $\tilde{\alpha}$ appears linearly, and
thus serves as a Lagrange multiplier. At this point, note since $\beta_i$ (or $\tilde{\beta}_i$) has already appeared quadratically, $\tilde{\alpha} + \lambda \beta_i^2$ with arbitrary constant $\lambda$, it is also a valid variable that acts as a Lagrange multiplier. This freedom will be used when matching the redefinition (\ref{alpha_mass_redef}) for the potential terms with that for the derivative terms [see the discussion around (\ref{cond_c_gen})].

On the other hand, the potential terms must always be accompanied with the
kinetic terms. If the kinetic term is taken to be GR, it is the lapse function $N$, or equivalently $\alpha$, that plays the role as a Lagrange multiplier and generates the corresponding (Hamiltonian) constraint.
The potential terms must be tuned to be compatible with this property. Thus,
(\ref{alpha_mass_redef}) implies another constraint
among parameters
	\begin{equation}
		\frac{1}{2}+\frac{3c'_{1}+c'_{2}}{b'_{1}}= 0 ,\label{c_cons2}
	\end{equation}
which together with (\ref{c_mass_cons1}) yields
	\[
		c'_{2}=-\frac{1}{2}b'_1 - 3c'_{1},\qquad c'_{3}=\frac{1}{2}b'_1+2c'_{1},
	\]
with a single free parameter $c'_{1}$ left undetermined. 
Finally, up to the cubic order, the potential terms compatible with GR are
	\begin{eqnarray}
		\mathcal{L}_{2}^{\mathrm{pot}}+\mathcal{L}_{3}^{\mathrm{pot}} & = & b'_1\left[\left(h^2-h_{\mu\nu}h^{\mu\nu}\right)+\frac{1}{2}\left(h^{3}-4h\, h_{\mu\nu}h^{\mu\nu}+3h_{\nu}^{\mu}h_{\rho}^{\nu}h_{\mu}^{\rho}\right)\right]\nonumber \\
		 &  & +\left(c'_{1}-\frac{1}{2}b'_1\right)\left(h^{3}-3h\, h_{\mu\nu}h^{\mu\nu}+2h_{\nu}^{\mu}h_{\rho}^{\nu}h_{\mu}^{\rho}\right),\label{L2L3_mass_fin}
	\end{eqnarray}
where $h\equiv h^{\mu}_{\mu}$.
It is easy to check that the first  line in (\ref{L2L3_mass_fin}) is nothing but the expansions of the first dRGT mass term
	\begin{equation}
		\mathcal{L}^{\mathrm{dRGT,1}} \equiv \sqrt{-g}\left(\left[\mathcal{K}\right]^{2}-\left[\mathcal{K}^{2}\right]\right),\label{L_dRGT_1}
	\end{equation}
up to the third order in $h_{\mu\nu}$ around a flat background in unitary gauge, while the second line in (\ref{L2L3_mass_fin}) corresponds to the expansion of 
	\begin{equation}
		\mathcal{L}^{\mathrm{dRGT,2}} \equiv \sqrt{-g}\left(\left[\mathcal{K}\right]^{3}-3\left[\mathcal{K}\right]\left[\mathcal{K}^{2}\right]+2\left[\mathcal{K}^{3}\right]\right). \label{L_dRGT_2}
	\end{equation}
Following this perturbative approach, one may in principle rearrive at the nonlinear structure (or the nonlinear completion) for the dRGT mass terms found in \cite{deRham:2010ik,deRham:2010kj}.

At this point, we emphasize again that the existence of the constraint (\ref{c_cons2}) and thus (\ref{L2L3_mass_fin}) are determined under the assumption that the kinetic term is given exactly by GR. 
In other words, the kinetic terms are given first, while the potential terms are added and tuned elaborately to be compatible with the constraint structure of the kinetic terms.
There is a possibility, however, that different potential terms (from the dRGT ones) may exist, but are compatible with different (non-GR) derivative terms.

\section{Derivative interactions at the cubic order}\label{sec:der}

We start from the most general Lorentz invariant cubic Lagrangian for $h_{\mu\nu}$ with two spacetime derivatives, which takes the form\footnote{Actually there are 16 independent contractions, among which 2 can be removed by linear combinations of other terms up to total derivatives. Thus here we choose the residual 14 terms the same as in \cite{deRham:2013tfa} (with different orders).}
	\begin{eqnarray}
	\mathcal{L}_{3}^{\mathrm{der}} & = & c_{1}h^{\alpha\beta}\partial_{\alpha}h^{\lambda\mu}\partial_{\beta}h_{\lambda\mu}+c_{2}h^{\alpha\beta}\partial_{\alpha}h_{\lambda}^{\lambda}\partial_{\beta}h_{\mu}^{\mu}+c_{3}h^{\alpha\beta}\partial_{\beta}h_{\alpha}^{\lambda}\partial_{\lambda}h_{\mu}^{\mu}+c_{4}h^{\alpha\beta}\partial_{\lambda}h_{\mu}^{\mu}\partial^{\lambda}h_{\alpha\beta}\nonumber \\
	 &  & +c_{5}h_{\alpha}^{\alpha}\partial_{\lambda}h_{\mu}^{\mu}\partial^{\lambda}h_{\beta}^{\beta}+c_{6}h^{\alpha\beta}\partial_{\lambda}h_{\alpha}^{\lambda}\partial_{\mu}h_{\beta}^{\mu}+c_{7}h^{\alpha\beta}\partial_{\beta}h_{\alpha}^{\lambda}\partial_{\mu}h_{\lambda}^{\mu}+c_{8}h_{\alpha}^{\alpha}\partial_{\beta}h^{\beta\lambda}\partial_{\mu}h_{\lambda}^{\mu}\nonumber \\
	 &  & +c_{9}h^{\alpha\beta}\partial^{\lambda}h_{\alpha\beta}\partial_{\mu}h_{\lambda}^{\mu}+c_{10}h_{\alpha}^{\alpha}\partial^{\lambda}h_{\beta}^{\beta}\partial_{\mu}h_{\lambda}^{\mu}+c_{11}h^{\alpha\beta}\partial_{\lambda}h_{\beta\mu}\partial^{\mu}h_{\alpha}^{\lambda}+c_{12}h^{\alpha\beta}\partial_{\mu}h_{\beta\lambda}\partial^{\mu}h_{\alpha}^{\lambda}\nonumber \\
	 &  & +c_{13}h_{\alpha}^{\alpha}\partial_{\lambda}h_{\beta\mu}\partial^{\mu}h^{\beta\lambda}+c_{14}h_{\alpha}^{\alpha}\partial_{\mu}h_{\beta\lambda}\partial^{\mu}h^{\beta\lambda},\label{L3_expl}
	\end{eqnarray}
where $c_1,\dots,c_{14}$ are constant parameters to be determined.

Together with $\mathcal{L}_2^{\mathrm{der}}$ given in (\ref{L2_EH}), after a tedious calculation, the expansion of $\mathcal{L}_2^{\mathrm{der}}+\mathcal{L}_3^{\mathrm{der}}$ in terms of $\{\alpha,\beta_i,h_{ij}\}$ at the cubic order can be schematically classified into 10 groups
	\begin{eqnarray}
	\mathcal{L}_{3}^{\mathrm{der,ADM}} & = & \mathcal{L}_{3,\alpha^{3}}+\mathcal{L}_{3,\alpha^{2}\beta}+\mathcal{L}_{3,\alpha\beta^{2}}+\mathcal{L}_{3,\beta^{3}}\nonumber \\
	 &  & +\mathcal{L}_{3,\alpha^{2}h}+\mathcal{L}_{3,\alpha\beta h}+\mathcal{L}_{3,\beta^{2}h}+\mathcal{L}_{3,\alpha h^{2}}+\mathcal{L}_{3,\beta h^{2}}+\mathcal{L}_{3,h^{3}},\label{L3_ADM_group}
	\end{eqnarray}
where the explicit expressions for each group of terms are given in Appendix \ref{sec:L3_ADM_expl}.
Requiring that $\mathcal{L}_3^{\mathrm{ADM}}$ can be put into the first order form with no time derivatives of $\alpha$ and $\beta_i$ yields 12 constraints [see (\ref{L3_cons_1})--(\ref{L3_cons_12})] on the 14 parameter $c_i$'s, from which we solve
	\begin{eqnarray}
	c_{2} & = & -c_{1},\qquad c_{3}=4c_{1},\qquad c_{4}=-2c_{1},\qquad c_{6}=-5c_{1}+2c_{5},\nonumber \\
	c_{7} & = & -4c_{1},\qquad c_{8}=-3c_{1}+2c_{5},\qquad c_{9}=2c_{1},\qquad c_{10}=-2c_{5},\label{c_sol_nd}\\
	c_{11} & = & 3c_{1}-2c_{5},\qquad c_{12}=2c_{1},\qquad c_{13}=3c_{1},\qquad c_{14}=-c_{5},\nonumber 
	\end{eqnarray}
with $c_1$ and $c_5$ left undetermined.

\subsection{Cubic Hamiltonian $\mathcal{H}_3$}

After plugging the solutions for parameters (\ref{c_sol_nd}), the cubic Lagrangian (\ref{L3_ADM_group}) can be written as
	\begin{equation}
	\mathcal{L}_{3}^{\mathrm{der,ADM}}=\frac{1}{2}\mathcal{G}_{ij,kl}^{(1)}\dot{h}_{ij}\dot{h}_{kl}+\mathcal{F}_{ij}^{(2)}\dot{h}_{ij}+\mathcal{W}^{(3)},\label{L3_ADM_expl}
	\end{equation}
where
	\begin{eqnarray}
	\mathcal{G}_{ij,kl}^{(1)} & = & \left(2\left(c_{1}-c_{5}\right)\alpha-c_{5}h_{mm}\right)\left(2\delta_{ij}\delta_{kl}-\delta_{ik}\delta_{jl}-\delta_{il}\delta_{jk}\right)\nonumber \\
	 &  & +c_{1}\left(2h_{ij}\delta_{kl}+2h_{kl}\delta_{ij}-h_{il}\delta_{jk}-h_{jl}\delta_{ik}-h_{ik}\delta_{jl}-h_{jk}\delta_{il}\right),\label{G1}
	\end{eqnarray}
and $\mathcal{F}_{ij}^{(2)}$ and $\mathcal{W}^{(3)}$ are given in \ref{sec:F2W3}.

Finally, together with the quadratic Lagrangian (\ref{L2_ADM_expl}), the Hamiltonian density at the cubic order can be written as
	\begin{equation}
	\mathcal{H}_{3}^{\mathrm{der}} =\mathcal{H}_{3,\alpha\pi^{2}}+\mathcal{H}_{3,\alpha^{2}h}+\mathcal{H}_{3,\beta\pi h} +\mathcal{H}_{3,\beta^{2}h} +\mathcal{H}_{3,h\pi^{2}} -\mathcal{W}_{\alpha h^{2}}^{(3)}-\mathcal{W}_{h^{3}}^{(3)},\label{H3_fin}
	\end{equation}
where
	\begin{eqnarray}
	\mathcal{H}_{3,\alpha\pi^{2}} & = & -\frac{1}{4b_{1}^{2}}\left(c_{1}-c_{5}\right)\alpha\left(\pi_{ii}\pi_{jj}-2\pi_{ij}\pi_{ij}\right),\label{H3_app}\\
	\mathcal{H}_{3,\alpha^{2}h} & = & 2\left(b_{1}+2\left(c_{1}-c_{5}\right)\right)\alpha^{2}\left(\partial_{i}\partial_{j}h_{ij}-\partial^2h_{ii}\right),\label{H3_aah}\\
	\mathcal{H}_{3,\beta\pi h} & = & \frac{1}{b_{1}}\beta_{i}\Big[\left(c_{1}-2c_{5}\right)\left(\pi_{ij}\left(\partial_{j}h_{kk}-\partial_{k}h_{jk}\right)-\frac{1}{2}\pi_{jj}\left(\partial_{i}h_{kk}+\partial_{k}h_{ik}\right)\right)\nonumber \\
	 &  & \qquad-c_{1}\pi_{jk}\partial_{i}h_{jk}+\left(3c_{1}-2c_{5}\right)\pi_{jk}\partial_{k}h_{ij}\Big],\label{H3_bph}
	\end{eqnarray}
	\begin{eqnarray}
	\mathcal{H}_{3,\beta^{2}h} & = & \left(c_{1}-2c_{5}\right)\Big[h_{ii}\left(\partial_{j}\beta_{j}\partial_{k}\beta_{k}+\partial_{j}\beta_{k}\partial_{k}\beta_{j}+2\beta_{j}\partial_{k}\partial_{j}\beta_{k}\right)\nonumber \\
	 &  & +2h_{ij}\left(\beta_{i}\partial^{2}\beta_{j}-\beta_{i}\partial_{k}\partial_{j}\beta_{k}-\beta_{k}\partial_{i}\partial_{k}\beta_{j}-\partial_{j}\beta_{i}\partial_{k}\beta_{k}-\partial_{j}\beta_{k}\partial_{k}\beta_{i}+\partial_{k}\beta_{j}\partial_{k}\beta_{i}\right)\Big]\nonumber \\
	 &  & +\left(2b_{1}+3c_{1}-2c_{5}\right)
	 \beta_{k}^2\left(\partial^{2}h_{ii}-\partial_{i}\partial_{j}h_{ij}\right),\label{H3_bbh}
	\end{eqnarray}
	\begin{equation}
	\mathcal{H}_{3,h\pi^{2}}=\frac{1}{8b_{1}^{2}}\left[c_{5}h_{ii}\left(\pi_{jj}\pi_{kk}-2\pi_{jk}\pi_{jk}\right)-2c_{1}h_{ij}\left(\pi_{ij}\pi_{kk}-2\pi_{ik}\pi_{jk}\right)\right],\label{H3_hpp}
	\end{equation}
and $\mathcal{W}_{\alpha h^{2}}^{(3)}$ and $\mathcal{W}_{h^{3}}^{(3)}$ are given in (\ref{W3_ahh}) and (\ref{W3_hhh}).

At the cubic order, because of the presence of $\mathcal{H}_{3,\alpha^2h}$ and $\mathcal{H}_{3,\beta^2h}$, both $\alpha$ and $\beta_i$ appear quadratically in the Hamiltonian. However, $\alpha^2$ and $\beta_i^2$ enter the Hamiltonian through a special combination [see (\ref{H3_aah}) and the last line of (\ref{H3_bbh})]
	\begin{equation}
		\mathcal{H}_{2}+\mathcal{H}_{3}\supset4b_{1}\left[\alpha+\left(\frac{1}{2}+\frac{c_{1}-c_{5}}{b_{1}}\right)\alpha^{2} - \frac{1}{2}\left(1+\frac{3c_{1}-2c_{5}}{2b_{1}}\right)\beta_{k}^{2}\right]\left(\partial_{i}\partial_{j}h_{ij}-\partial^{2}h_{ii}\right),
	\end{equation}
where we also include the relevant term from $\mathcal{H}_2^{\mathrm{der}}$.
Thus, we can always introduce a new variable\footnote{Generally, as the case of (\ref{alpha_mass_redef}), we may write $\tilde{\alpha}\propto\alpha+\left(\frac{1}{2}+\frac{c_{1}-c_{5}}{b_{1}}\right)\alpha^{2} +\lambda \beta_{k}^{2}$ with arbitrary numerical constant $\lambda$ such that $\tilde{\alpha}$ enters the Hamiltonian linearly. In (\ref{alpha_redef}), we fix $\lambda = -\frac{1}{2}\left(1+\frac{3c_{1}-2c_{5}}{2b_{1}}\right)$, since with this choice of $\tilde{\alpha}$, $\beta_i$ explicitly appears linearly in the Hamiltonian when $c_1 = 2c_5$ (while with other values of $\lambda$, this degeneration is not transparent), which corresponds to the GR case.}
	\begin{equation}
	\tilde{\alpha}\propto\alpha+\left(\frac{1}{2}+\frac{c_{1}-c_{5}}{b_{1}}\right)\alpha^{2}- \frac{1}{2}\left(1+\frac{3c_{1}-2c_{5}}{2b_{1}}\right)\beta_{k}^{2},\label{alpha_redef}
	\end{equation}
such that, up to cubic order, $\tilde{\alpha}$ appears linearly in the Hamiltonian.	This also implies that there is no further constraint between $c_1$ and $c_5$; in other words, simply requiring the vanishing of $\dot{\alpha}$ and $\dot{\beta}_i$ has already been sufficient to determine the derivative terms.

\subsection{Derivative terms and the compatible potential terms}

Up to the cubic order in $h_{\mu\nu}$, we are left with a set of (generally nondiffeomorphism invariant) derivative terms with two parameters, which can be written in the form\footnote{More explicitly, in terms of (\ref{L3_expl}), $\mathcal{L}^{\mathrm{GR}}_3$ corresponds to the choice of parameters  with (\ref{c_sol_nd}) and $c_1 = 2c_5=1/4$, while $\mathcal{L}^{\mathrm{PL}}$ corresponds to (\ref{c_sol_nd}) and $c_1 = c_5 = 1$.}
	\begin{equation}
		\mathcal{L}_2^{\mathrm{der}}+\mathcal{L}_3^{\mathrm{der}} \simeq -4b_{1}\left(\mathcal{L}_{2}^{\mathrm{GR}}+2\frac{c_{5}-c_{1}}{b_{1}}\mathcal{L}_{3}^{\mathrm{GR}}+\frac{c_{1}-2c_{5}}{4b_{1}}\mathcal{L}^{\mathrm{PL}}\right), \label{L_der_fin}
	\end{equation}
where $\mathcal{L}^{\mathrm{GR}}$ denotes $\sqrt{-g}R$, the subscripts denote the expansions in $h_{\mu\nu}$ around the Minkowski background at the quadratic/cubic order,  respectively, $\mathcal{L}^{\mathrm{PL}}$ stands for the so-called ``pseudolinear derivative term'' (antisymmetrization is unnormalized) 
	\begin{equation}
		\mathcal{L}^{\mathrm{PL}}\simeq \eta_{\nu_{1}}^{[\mu_{1}}\eta_{\nu_{2}}^{\mu_{2}}\eta_{\nu_{3}}^{\mu_{3}}\eta_{\nu_{4}}^{\mu_{4}]}h_{\mu_{1}}^{\nu_{1}}\partial^{\nu_{2}}h_{\mu_{3}}^{\nu_{3}}\partial_{\mu_{2}}h_{\mu_{4}}^{\nu_{4}}, \label{L_PL_compa}
	\end{equation}
which was identified in \cite{Hinterbichler:2013eza} (see also \cite{Folkerts:2011ev,Folkerts:2013mra}).

Note in the case of no pseudolinear derivative terms, i.e. $c_1 = 2c_5$, Eq. (\ref{L_der_fin}) seems  different from the standard expansion of GR if $c_{5}-c_{1}\neq b_{1}/2$. However, this apparent discrepancy can be trivially removed by a field rescaling
	\begin{equation}
		h_{\mu\nu}=\frac{b_{1}}{2\left(c_{5}-c_{1}\right)}\tilde{h}_{\mu\nu}, \label{h_rescale}
	\end{equation}
when $c_1 \neq c_5$.
The derivative terms can be recast in terms of $\tilde{h}_{\mu\nu}$ as
	\begin{equation}
		\mathcal{L}_{2}^{\mathrm{der}}+\mathcal{L}_{3}^{\mathrm{der}}=-\frac{b_{1}^{3}}{\left(c_{5}-c_{1}\right)^{2}}\left(\mathcal{L}_{2}^{\mathrm{GR}}[\tilde{h}]+\mathcal{L}_{3}^{\mathrm{GR}}[\tilde{h}]+\frac{c_{1}-2c_{5}}{8\left(c_{5}-c_{1}\right)}\mathcal{L}^{\mathrm{PL}}[\tilde{h}]\right).\label{L_der_rescale}
	\end{equation}
With (\ref{L_der_rescale}), it becomes transparent that, up to the cubic order, the allowed derivative terms are simply the linear combination of the standard GR terms and the pseudolinear derivative term.
GR is recovered (perturbatively) for $c_1 = 2c_5$, while the pseudolinear derivative term is recovered for $c_1 = c_5$ [where the Lagrangian is given through (\ref{L_der_fin})].

As we have discussed in Sec.\ref{sec:mass_term}, the diffeomorphism-breaking derivative terms must be accompanied with the appropriate potential terms. Comparing (\ref{alpha_redef}) with (\ref{alpha_mass_redef}), in order to make these two redefinitions for $\tilde{\alpha}$ consistent, we must have
	\begin{equation}
		\frac{3c_{1}'+c_{2}'}{b_{1}'}=\frac{c_{1}-c_{5}}{b_{1}}, \label{cond_c_gen}
	\end{equation}
which yield a constraint among $c'_1$, $c'_2$, which are parameters for the potential terms, and $c_1$, $c_5$, which are parameters for the derivative terms. 
Note (\ref{cond_c_gen}) only involves coefficients in front of $\alpha^2$ terms in (\ref{alpha_redef}) and (\ref{alpha_mass_redef}), since $\beta_i^2$ terms in both case can be tuned freely [although we have fixed the coefficient of the $\beta_i^2$ term in (\ref{alpha_redef})] and thus can be matched.
Moreover, we have the same redefinition for $\tilde{\beta}_i$ (\ref{beta_mass_redef}) in order to make $\tilde{\alpha}$ appears linearly in the Hamiltonian coming from the potential terms.
Finally, after some manipulations, the potential terms can be written as
	\begin{equation}
		\mathcal{L}_{2}^{\mathrm{pot}}+ \mathcal{L}_{3}^{\mathrm{pot}}=4b_{1}'\left[\mathcal{L}_{2}^{\mathrm{dRGT,1}}+2\frac{c_{5}-c_{1}}{b_{1}}\mathcal{L}_{3}^{\mathrm{dRGT,1}}+2\left(\frac{c_{1}'}{b_{1}'}+\frac{c_{1}-c_{5}}{b_{1}}\right)\mathcal{L}_{3}^{\mathrm{dRGT,2}}\right], \label{L_pot_fin}
	\end{equation}
where $\mathcal{L}^{\mathrm{dRGT},1}$ and $\mathcal{L}^{\mathrm{dRGT},2}$ are the dRGT nonlinear mass terms given in (\ref{L_dRGT_1}) and (\ref{L_dRGT_2}), respectively, and again the subscripts denote the expansions at the corresponding orders.
Similar to the derivative terms, for $c_{5}\neq c_{1}$, Eq. (\ref{L_pot_fin}) can be recast in terms of the same rescaled $\tilde{h}_{\mu\nu}$ as (\ref{h_rescale}),
	\begin{equation}
		\mathcal{L}_{2}^{\mathrm{pot}}+ \mathcal{L}_{3}^{\mathrm{pot}} = \frac{b_{1}'b_{1}^{2}}{\left(c_{5}-c_{1}\right)^{2}}\left[\mathcal{L}_{2}^{\mathrm{dRGT,1}}[\tilde{h}]+\mathcal{L}_{3}^{\mathrm{dRGT,1}}[\tilde{h}]+\left(1+\frac{b_{1}c_{1}'}{b_{1}'\left(c_{5}-c_{1}\right)}\right)\mathcal{L}_{3}^{\mathrm{dRGT,2}}[\tilde{h}]\right], \label{L_pot_rescale}
	\end{equation}
which is explicitly the linear combination of the expansions of $\mathcal{L}_{3}^{\mathrm{dRGT,1}}$ and $\mathcal{L}_{3}^{\mathrm{dRGT,2}}$.

At this point, we can understand how GR terms and the pseudolinear derivative term (\ref{L_PL_compa}) arise  with different choices of parameters:
\begin{enumerate}[(i)]
\item GR terms

According to (\ref{L_der_fin}) or equivalently (\ref{L_der_rescale}), the derivative terms reduce to the GR form when $c_1 = 2c_5$. In this case, Eq. (\ref{alpha_redef}) becomes
	\begin{equation}
		\tilde{\alpha} = \alpha+\frac{1}{2}\left(1+\frac{c_{1}}{b_{1}}\right)\left(\alpha^{2}-\beta_{i}^{2}\right),
	\end{equation}
which is nothing but the perturbation of lapse $\delta\tilde{N}$ corresponding to the rescaled $\tilde{h}_{\mu\nu}$ in (\ref{h_rescale}).\footnote{This can be verified directly by evaluating $\delta\tilde{N}=-\frac{1}{2}\tilde{h}_{00}-\frac{1}{8}\left(\tilde{h}_{00}^{2}-4\tilde{h}_{0i}\tilde{h}_{0i}\right)+\mathcal{O} (\tilde{h}^{3})$.} This is consistent with the well-known conclusion that in GR it is the lapse function that generates the Hamiltonian constraint. When there is no potential term, with the set of $\{\tilde{\alpha},\beta_i\}$, $\beta_i$ also appears linearly in the Hamiltonian [since the first two lines in (\ref{H3_bbh}) vanish] and thus generates three momentum constraints. 
When the potential terms are included, as has been discussed in Sec.\ref{sec:mass_term}, or more generally as in (\ref{L_pot_rescale}), only potential terms of the dRGT form are allowed.

\item Pseudolinear derivatives

According to (\ref{L_der_fin}), this corresponds to  $c_1=c_5 \neq 0$ and $c'_2 = -3c'_1 \neq 0$, and thus both sides of (\ref{cond_c_gen}) identically vanish.
In this case, the cubic derivative terms are fixed to be $\mathcal{L}^{\mathrm{PL}}$, and the cubic potential terms (\ref{L3_mass}) are fixed to be $c'_{1}\left(h^{3}-3h\, h_{\mu\nu}h^{\mu\nu}+2h_{\nu}^{\mu}h_{\rho}^{\nu}h_{\mu}^{\rho}\right) \equiv 8c_1' \mathcal{L}_{3}^{\mathrm{dRGT,2}}$, which is the generalization of the quadratic Fierz-Pauli mass term to the cubic order, i.e., the cubic pseudolinear potential term. 
In this case, Eq. (\ref{alpha_redef}) reduces to
	\begin{equation}
		\tilde{\alpha} = \alpha+\frac{1}{2}\alpha^{2}-\frac{1}{2}\left(1+\frac{c_{1}}{2b_{1}}\right)\beta_{i}^{2} = -\frac{1}{2}h_{00}-\frac{c_{1}}{4b_{1}}h_{0i}^{2}.
	\end{equation}
Since now $\mathcal{H}_{3,\beta^2h}$ does not vanish, $\beta_i\equiv h_{0i}$ has already appeared quadratically in the Hamiltonian, and it is essentially $h_{00}$ that acts as the Lagrange multiplier.\footnote{In fact, the pseudo-linear terms in \cite{Hinterbichler:2013eza} are derived just based on the requirement that $h_{00}$ appears linearly in the Hamiltonian.} 
\end{enumerate}
To summarize, by combining (\ref{L_der_rescale}) and (\ref{L_pot_rescale}), up to the cubic order in the spin-2 field, the full Lorentz invariant kinetic and potential terms propagating at most 5 degrees of freedom can be conveniently written as 
	\begin{equation}
		\sim\mathcal{L}_{2}^{\mathrm{GR}}[h]+\mathcal{L}_{3}^{\mathrm{GR}}[h]+\lambda_{1}\mathcal{L}^{\mathrm{PL}}[h]+m^{2}\left(\mathcal{L}_{2}^{\mathrm{dRGT,1}}[h]+\mathcal{L}_{3}^{\mathrm{dRGT,1}}[h]+\lambda_{2}\mathcal{L}_{3}^{\mathrm{dRGT,2}}[h]\right), \label{fin_der_pot}
	\end{equation}
up to the overall factor, with three free parameters
	\[
		\lambda_{1}=\frac{c_{1}-2c_{5}}{8\left(c_{5}-c_{1}\right)},\qquad\lambda_{2}=1+\frac{b_{1}c_{1}'}{b_{1}'\left(c_{5}-c_{1}\right)},\qquad m^{2}=-\frac{b_{1}'}{b_{1}},
	\]
where in (\ref{fin_der_pot}) we have replaced $\tilde{h}_{\mu\nu}$ by $h_{\mu\nu}$.
Thus up to the cubic order, comparing with the dRGT massive gravity, there is an additional parameter $\lambda_1$ that characterizes the pseudolinear derivative terms.

\section{Conclusion}\label{sec:con}

In this article we studied the Lorentz invariant two-derivative interactions for a spin-2 field, up to the cubic order.
Through a perturbative ADM analysis, we determine the parameters by requiring the existence of a combination of perturbation of lapse $\alpha\equiv \delta N$ and shift $\beta_i$, which enters the Hamiltonian linearly and thus generates a constraint that may eliminate the ghost degree of freedom. 
Simply demanding that $\{\alpha,\beta_i\}$ have vanishing conjugate momenta yields the set of cubic derivative terms (\ref{L_der_fin}) satisfying this requirement, which corresponds to the linear combination of the standard GR terms and the pseudolinear derivative term (\ref{L_PL_compa}) first identified in \cite{Hinterbichler:2013eza}. 
The resulting derivative terms possess a Lagrange multiplier given in (\ref{alpha_redef}), which is a nonlinear function of $\{\alpha,\beta_i\}$ generally and is responsible for generating the constraint that removes the ghost. 
After fixing the derivative terms, the potential terms should be included without violating (\ref{alpha_redef}), which yields an additional constraint on parameters (\ref{cond_c_gen}), and the resulting set of potential terms corresponds to the linear combination of the first two dRGT mass terms.

Our results confirm the existence of the pseudolinear derivative term at cubic order. It is thus interesting to extend our formalism to higher orders and to see if there exist non-GR derivative terms at higher orders.
However, according to the ``no-go'' theorem for such non-GR derivative terms \cite{deRham:2013tfa}, the cubic pseudolinear derivative term is not the leading expansion of any ghost-free nonlinear terms. This prevents it from being a viable theory of massive gravity, since any coupling with matter will inevitably push the theory toward a fully nonlinear level. 
On the other hand, this no-go theorem makes not only GR but also the pseudolinear derivative term itself very special, which deserves further investigations.
The lack of nonlinear completion for the pseudolinear derivative term, is reminiscent of the case of vector field(s). For example, a single vector field with $U(1)$ gauge symmetry does not have self-interactions on a flat background \cite{Deffayet:2013tca}, while (nonlinear) derivative interactions are allowed when there are multiple vector fields  \cite{Deffayet:2010zh} or the gauge symmetry is abandoned \cite{Tasinato:2014eka,Heisenberg:2014rta}. Thus, there is a  possibility that in the bi/multimetric framework, or Lorentz-breaking context (e.g. \cite{Lin:2013aha,Lin:2013sja}), ghost-free non-GR derivative interactions may exist.

\acknowledgments

I would like to thank Masahide Yamaguchi for interesting discussions.
I was supported by JSPS Grant-in-Aid for Scientific Research No. 25287054.

\appendix
\section{Linearized Einstein-Hilbert}\label{sec:linear_EH}

To illuminate the basic procedure in this work, here we briefly show how the linearized Einstein-Hilbert Lagrangian is uniquely determined by the constraint analysis.
See \cite{VanNieuwenhuizen:1973fi} for a general discussion on the dependence of the number of d.o.f. on the kinetic terms.

Consider a general Lorentz invariant kinetic term for metric perturbation $h_{\mu\nu}$,
	\begin{equation}\label{L2_gen}
		\mathcal{L}_2 = \Gamma_{2}\, \partial_{\alpha}h_{\mu\nu}\partial_{\beta}h_{\rho\sigma},
	\end{equation}
where the ``coefficient'' $\Gamma_{2}$ is a trinomial of the Minkowski metric $\eta^{\mu\nu}$, which takes the following general form:
	\begin{equation}\label{Gamma_2_def}
		\Gamma_{2} = b_{1}\,\eta^{\alpha\beta}\eta^{\mu\rho}\eta^{\nu\sigma}+b_{2}\,\eta^{\alpha\rho}\eta^{\beta\sigma}\eta^{\mu\nu}+b_{3}\,\eta^{\alpha\nu}\eta^{\beta\sigma}\eta^{\mu\rho}+b_{4}\,\eta^{\alpha\beta}\eta^{\mu\nu}\eta^{\rho\sigma},
	\end{equation}
with four undetermined numerical coefficients $b_1,\dots,b_4$. In terms of the perturbative ADM variables $\{\alpha,\beta_i,h_{ij}\}$, at the quadratic order and up to total derivatives, Eq. (\ref{L2_gen}) yields
	\begin{equation}\label{L2_ADM}
		\mathcal{L}_{2}^{\mathrm{ADM}}=\mathcal{L}_{2,\alpha^{2}}+\mathcal{L}_{2,\alpha\beta}+\mathcal{L}_{2,\beta^{2}}+\mathcal{L}_{2,\alpha h}+\mathcal{L}_{2,\beta h}+\mathcal{L}_{2,h^{2}},
	\end{equation}
with
	\begin{eqnarray}
	\mathcal{L}_{2,\alpha^{2}} & = & -4\left(b_{1}+b_{2}+b_{3}+b_{4}\right)\dot{\alpha}^{2}+4\left(b_{1}+b_{4}\right)\partial_{i}\alpha\partial_{i}\alpha,\label{L2_aa}\\
	\mathcal{L}_{2,\alpha\beta} & = & -4\left(b_{2}+b_{3}\right)\dot{\alpha}\partial_{i}\beta_{i},\label{L2_ab}\\
	\mathcal{L}_{2,\beta^{2}} & = & \left(2b_{1}+b_{3}\right)\dot{\beta}_{i}\dot{\beta}_{i}-b_{3}\partial_{i}\beta_{i}\partial_{j}\beta_{j}-2b_{1}\partial_{j}\beta_{i}\partial_{j}\beta_{i},\label{L2_bb}\\
	\mathcal{L}_{2,\alpha h} & = & -2\left(b_{2}+2b_{4}\right)\dot{\alpha}\dot{h}_{ii}+2\partial_{j}\alpha\left(b_{2}\partial_{i}h_{ij}+2b_{4}\partial_{j}h_{ii}\right),\label{L2_at}\\
	\mathcal{L}_{2,\beta h} & = & -2b_{2}\partial_{i}\beta_{i}\dot{h}_{jj}-2b_{3}\partial_{j}\beta_{i}\dot{h}_{ij},,\label{L2_bt}\\
	\mathcal{L}_{2,h^{2}} & = & -b_{1}\dot{h}_{ij}\dot{h}_{ij}-b_{4}\dot{h}_{ii}\dot{h}_{jj}\nonumber \\
	 &  & +b_{1}\partial_{k}h_{ij}\partial_{k}h_{ij}+b_{2}\partial_{j}h_{ii}\partial_{k}h_{kj}+b_{3}\partial_{j}h_{ij}\partial_{k}h_{ik}+b_{4}\partial_{j}h_{ii}\partial_{j}h_{kk}.\label{L2_tt}
	\end{eqnarray}
To prevent $\alpha$ and $\beta_i$ from being dynamical, terms in the above proportional to $\dot{\alpha}^{2}$, $\dot{\alpha}\partial_{i}\beta_{i}$, $\dot{\beta}_{i}\dot{\beta}_{i}$, and $\dot{\alpha}\dot{h}_{ii}$ must be identically vanishing, which implies
	\[
		b_{1}+b_{2}+b_{3}+b_{4}=2b_{1}+b_{3}=b_{2}+b_{3}=b_{2}+2b_{4}=0,
	\]
which \emph{uniquely} fixes the four coefficients up to an overall factor
	\begin{equation}
		b_{2}=2b_{1},\qquad b_{3}=-2b_{1},\qquad b_{4}=-b_{1}.
	\end{equation}
It is interesting to see that with these coefficients, $\partial_i\alpha \partial_i \alpha$ also drops out from the Lagrangian, which makes it as a Lagrange multiplier.
The final quadratic Lagrangian is thus given by (\ref{L2_EH}), which is just the linearization of the Einstein-Hilbert term.

\section{Explicit expressions}

Here we collect some explicit expressions, whose quantitative forms are needed in our discussion.

\subsection{$\mathcal{L}_3^{\mathrm{der,ADM}}$}\label{sec:L3_ADM_expl}

Up to total derivatives, we have
\begin{equation}
\mathcal{L}_{3,\alpha^{3}}=-8\left(\sum_{i=1}^{14}c_{i}\right)\alpha\dot{\alpha}^{2}+8\left(c_{4}+c_{5}+c_{12}+c_{14}\right)\alpha\partial_{i}\alpha\partial_{i}\alpha,\label{L3_aaa}
\end{equation}
\begin{eqnarray}
\mathcal{L}_{3,\alpha^{2}\beta} & = & -4\left(c_{3}+2c_{6}+c_{7}+2c_{8}+2c_{9}+2c_{10}+2c_{11}+2c_{13}\right)\alpha\dot{\alpha}\partial_{i}\beta_{i}\nonumber \\
 &  & -4\left(2c_{1}+2c_{2}+c_{3}+c_{7}\right)\dot{\alpha}\partial_{i}\alpha\beta_{i}.\label{L3_aab}
\end{eqnarray}
\begin{eqnarray}
\mathcal{L}_{3,\alpha\beta^{2}} & = & 2\left(2c_{1}+c_{7}+c_{8}+c_{12}+c_{13}+2c_{14}\right)\alpha\dot{\beta}_{i}\dot{\beta}_{i}-2\left(c_{3}+c_{6}+c_{8}\right)\alpha\partial_{i}\beta_{i}\partial_{j}\beta_{j}\nonumber \\
 &  & +2\left(c_{3}+2c_{4}+2c_{6}+c_{7}+2c_{9}+2c_{11}+2c_{12}\right)\dot{\alpha}\beta_{i}\dot{\beta}_{i}+2\left(c_{3}-c_{11}-c_{13}\right)\alpha\partial_{i}\beta_{j}\partial_{j}\beta_{i}\nonumber \\
 &  & -2\left(c_{12}+2c_{14}\right)\alpha\partial_{i}\beta_{j}\partial_{i}\beta_{j}-2\left(c_{3}+c_{7}\right)\partial_{i}\alpha\beta_{i}\partial_{j}\beta_{j}-4\left(c_{4}+c_{12}\right)\partial_{i}\alpha\beta_{j}\partial_{i}\beta_{j},\label{L3_abb}
\end{eqnarray}
\begin{eqnarray}
\mathcal{L}_{3,\beta^{3}} & = & \left(2c_{6}+c_{7}+4c_{9}+2c_{11}\right)\beta_{i}\dot{\beta}_{i}\partial_{j}\beta_{j}+\left(4c_{1}+c_{7}\right)\beta_{i}\dot{\beta}_{j}\partial_{i}\beta_{j}.\label{L3_bbb}
\end{eqnarray}
\begin{eqnarray}
\mathcal{L}_{3,\alpha^{2}h}&=&-4\left(c_{5}+c_{8}+c_{10}+c_{13}+c_{14}\right)\dot{\alpha}^{2}h_{ii}-4\left(2c_{2}+c_{3}+c_{4}+2c_{5}+c_{10}\right)\alpha\dot{\alpha}\dot{h}_{ii}\nonumber\\&&+4\left(c_{5}+c_{14}\right)\partial_{i}\alpha\partial_{i}\alpha h_{jj}+4\left(c_{1}+c_{2}\right)\partial_{i}\alpha\partial_{j}\alpha h_{ij}\nonumber\\&&+2\left(b_{1}-c_{4}-2c_{5}\right)\alpha^{2}\partial^{2}h_{ii}-2\left(b_{1}+c_{9}+c_{10}\right)\alpha^{2}\partial_{i}\partial_{j}h_{ij}.\label{L3_aah}
\end{eqnarray}
\begin{eqnarray}
\mathcal{L}_{3,\alpha\beta h} & = & -4\left(c_{8}+c_{10}+c_{13}\right)\dot{\alpha}\partial_{i}\beta_{i}h_{jj}-2\left(c_{3}+c_{7}\right)\dot{\alpha}\partial_{i}\beta_{j}h_{ij}-2\left(c_{3}+2c_{6}+2c_{11}\right)\dot{\alpha}\beta_{i}\partial_{j}h_{ij}\nonumber \\
 &  & -2\left(c_{3}+c_{7}+2c_{8}+2c_{11}\right)\alpha\dot{\beta}_{i}\partial_{j}h_{ij}-2\left(4c_{2}+c_{3}\right)\dot{\alpha}\beta_{i}\partial_{i}h_{jj}-4\left(c_{2}+c_{3}-c_{13}\right)\alpha\dot{\beta}_{i}\partial_{i}h_{jj}\nonumber \\
 &  & +2\left(2c_{2}+c_{3}-2c_{10}-2c_{13}\right)\alpha\partial_{i}\beta_{i}\dot{h}_{jj}+2\left(c_{3}+2c_{11}-2c_{13}\right)\alpha\partial_{i}\beta_{j}\dot{h}_{ij}.\label{L3_abh}
\end{eqnarray}
\begin{eqnarray}
\mathcal{L}_{3,\beta^{2}h} & = & \left(c_{6}+c_{11}+c_{12}\right)\dot{\beta}_{i}\dot{\beta}_{j}h_{ij}+\left(c_{8}+c_{13}+2c_{14}\right)\dot{\beta}_{i}\dot{\beta}_{i}h_{jj}+\left(c_{7}+2c_{12}\right)\beta_{i}\dot{\beta}_{j}\dot{h}_{ij}\nonumber \\
 &  & +\left(c_{3}+2c_{4}\right)\beta_{i}\dot{\beta}_{i}\dot{h}_{jj}-c_{8}\partial_{i}\beta_{i}\partial_{j}\beta_{j}h_{kk}-c_{13}\partial_{i}\beta_{j}\partial_{j}\beta_{i}h_{kk}-2c_{14}\partial_{i}\beta_{j}\partial_{i}\beta_{j}h_{kk}\nonumber \\
 &  & -2c_{1}\partial_{i}\beta_{j}\partial_{k}\beta_{j}h_{ik}-c_{7}\partial_{i}\beta_{j}\partial_{k}\beta_{k}h_{ij}-c_{12}\partial_{i}\beta_{j}\partial_{i}\beta_{k}h_{jk}-c_{3}\beta_{i}\partial_{i}\beta_{j}\partial_{j}h_{kk}\nonumber \\
 &  & -c_{7}\beta_{i}\partial_{i}\beta_{j}\partial_{k}h_{jk}-2c_{11}\beta_{i}\partial_{k}\beta_{j}\partial_{j}h_{ik}+2\left(2b_{1}-c_{4}\right)\beta_{i}\partial_{j}\beta_{i}\partial_{j}h_{kk}-2c_{6}\beta_{i}\partial_{j}\beta_{j}\partial_{k}h_{ik}\nonumber \\
 &  & -2\left(2b_{1}+c_{9}\right)\beta_{i}\partial_{j}\beta_{i}\partial_{k}h_{jk}-2c_{12}\beta_{i}\partial_{j}\beta_{k}\partial_{j}h_{ik}.\label{L3_bbh}
\end{eqnarray}
\begin{eqnarray}
\mathcal{L}_{3,\alpha h^{2}} & = & -2\left(2c_{5}+c_{10}\right)\dot{\alpha}h_{ii}\dot{h}_{jj}-2\left(c_{4}+c_{9}\right)\dot{\alpha}h_{ij}\dot{h}_{ij}\nonumber \\
 &  & +\alpha\Big\{-2\left(c_{1}+c_{14}\right)\dot{h}_{ij}\dot{h}_{ij}-2\left(c_{2}+c_{5}\right)\dot{h}_{ii}\dot{h}_{jj}-4c_{5}\partial^{2}h_{jj}h_{kk}-2c_{4}\partial^{2}h_{jk}h_{jk}\nonumber \\
 &  & -2c_{10}\partial_{i}\partial_{j}h_{ij}h_{kk}-4c_{2}\partial_{i}\partial_{j}h_{kk}h_{ij}-2c_{3}\partial_{i}\partial_{j}h_{ik}h_{jk}-2c_{5}\partial_{i}h_{jj}\partial_{i}h_{kk}-4c_{2}\partial_{j}h_{ij}\partial_{i}h_{kk}\nonumber \\
 &  & -2\left(c_{4}-c_{14}\right)\partial_{i}h_{jk}\partial_{i}h_{jk}-2\left(c_{3}-c_{13}\right)\partial_{k}h_{ij}\partial_{i}h_{jk}+2c_{8}\partial_{i}h_{ij}\partial_{k}h_{jk}\Big\}.\label{L3_ahh}
\end{eqnarray}
\begin{eqnarray}
\mathcal{L}_{3,\beta h^{2}} & = & 2\left(c_{6}-c_{7}\right)\beta_{i}\dot{h}_{ij}\partial_{k}h_{jk}+\left(c_{7}-2c_{6}\right)\beta_{i}\partial_{k}h_{ij}\dot{h}_{jk}-c_{3}\partial_{j}\beta_{i}h_{ij}\dot{h}_{kk}+2c_{8}\beta_{i}\partial_{k}h_{ik}\dot{h}_{jj}\nonumber \\
 &  & -\left(c_{3}+2c_{8}\right)\beta_{i}\dot{h}_{ik}\partial_{k}h_{jj}-2\left(c_{6}+c_{11}\right)\partial_{k}\beta_{i}h_{ij}\dot{h}_{jk}-2c_{10}\partial_{i}\beta_{i}\dot{h}_{jj}h_{kk}\nonumber \\
 &  & -2\left(c_{8}+c_{13}\right)\partial_{k}\beta_{j}h_{ii}\dot{h}_{jk}-2c_{9}\partial_{k}\beta_{k}h_{ij}\dot{h}_{ij}-c_{7}\partial_{k}\beta_{i}\dot{h}_{ij}h_{jk}\nonumber \\
 &  & -2c_{2}\beta_{i}\partial_{i}h_{jj}\dot{h}_{kk}-2c_{1}\beta_{i}\partial_{i}h_{jk}\dot{h}_{jk}.\label{L3_bhh}
\end{eqnarray}
\begin{eqnarray}
\mathcal{L}_{3,h^{3}} & \equiv & -c_{4}h_{ij}\dot{h}_{ij}\dot{h}_{kk}-c_{5}h_{ii}\dot{h}_{jj}\dot{h}_{kk}-c_{12}h_{ij}\dot{h}_{ik}\dot{h}_{jk}-c_{14}h_{ii}\dot{h}_{jk}\dot{h}_{jk}+c_{1}h_{ij}\partial_{i}h_{kl}\partial_{j}h_{kl}\nonumber \\
 &  & +c_{2}h_{ij}\partial_{i}h_{kk}\partial_{j}h_{ll}+c_{3}h_{ij}\partial_{j}h_{ik}\partial_{k}h_{ll}+c_{4}h_{ij}\partial_{k}h_{ll}\partial_{k}h_{ij}+c_{5}h_{ii}\partial_{k}h_{ll}\partial_{k}h_{jj}\nonumber \\
 &  & +c_{6}h_{ij}\partial_{k}h_{ik}\partial_{l}h_{jl}+c_{7}h_{ij}\partial_{j}h_{ik}\partial_{l}h_{kl}+c_{8}h_{ii}\partial_{j}h_{jk}\partial_{l}h_{kl}+c_{9}h_{ij}\partial_{k}h_{ij}\partial_{l}h_{kl}\nonumber \\
 &  & +c_{10}h_{ii}\partial_{k}h_{jj}\partial_{l}h_{kl}+c_{11}h_{ij}\partial_{k}h_{jl}\partial_{l}h_{ik}+c_{12}h_{ij}\partial_{l}h_{jk}\partial_{l}h_{ik}\nonumber \\
 &  & +c_{13}h_{ii}\partial_{k}h_{jl}\partial_{l}h_{jk}+c_{14}h_{ii}\partial_{l}h_{jk}\partial_{l}h_{jk}.\label{L3_hhh}
\end{eqnarray}

We make many integrations by parts so that (\ref{L3_aaa})--(\ref{L3_hhh}) have  already been put into the first order form (in time derivatives). The vanishing of the time derivatives of $\alpha$ and $\beta_i$ yields 12 linearly independent algebraic equations:
	\begin{eqnarray}
	\sum_{i=1}^{14}c_{i} & = & 0,\label{L3_cons_1}\\
	c_{3}+2c_{6}+c_{7}+2c_{8}+2c_{9}+2c_{10}+2c_{11}+2c_{13} & = & 0,\label{L3_cons_2}\\
	2c_{1}+2c_{2}+c_{3}+c_{7} & = & 0,\label{L3_cons_3}\\
	2c_{1}+c_{7}+c_{8}+c_{12}+c_{13}+2c_{14} & = & 0,\label{L3_cons_4}\\
	2c_{6}+c_{7}+4c_{9}+2c_{11} & = & 0,\label{L3_cons_5}\\
	4c_{1}+c_{7} & = & 0,\label{L3_cons_6}\\
	2c_{2}+c_{3}+c_{4}+2c_{5}+c_{10} & = & 0,\label{L3_cons_7}\\
	c_{8}+c_{10}+c_{13} & = & 0,\label{L3_cons_8}\\
	c_{3}+c_{7}+2c_{8}+2c_{11} & = & 0,\label{L3_cons_9}\\
	4c_{2}+c_{3} & = & 0,\label{L3_cons_10}\\
	c_{2}+c_{3}-c_{13} & = & 0,\label{L3_cons_11}\\
	c_{6}+c_{11}+c_{12} & = & 0,\label{L3_cons_12}
	\end{eqnarray}
of which the solutions are given in (\ref{c_sol_nd}).

\subsection{$\mathcal{F}_{ij}^{(2)}$ and $\mathcal{W}^{(3)}$}\label{sec:F2W3}

	\begin{eqnarray}
	\mathcal{F}_{ij}^{(2)} & = & \left(-2c_{1}\partial_{k}\beta_{l}h_{kl}-\left(3c_{1}-2c_{5}\right)\beta_{k}\partial_{l}h_{kl}+2c_{5}\partial_{l}\beta_{l}h_{kk}+c_{1}\beta_{l}\partial_{l}h_{kk}\right)\delta_{ij}\nonumber \\
	 &  & -\left(c_{1}-2c_{5}\right)\beta_{i}\partial_{k}h_{jk}+\left(3c_{1}-2c_{5}\right)\beta_{k}\partial_{i}h_{kj}+\left(c_{1}-2c_{5}\right)\beta_{i}\partial_{j}h_{kk}\nonumber \\
	 &  & +2c_{1}\partial_{i}\beta_{k}h_{kj}-2c_{5}\partial_{i}\beta_{j}h_{kk}+2c_{1}\partial_{k}\beta_{i}h_{jk}-2c_{1}\partial_{k}\beta_{k}h_{ij}-c_{1}\beta_{k}\partial_{k}h_{ij}\nonumber \\
	 &  & +4\left(c_{1}-c_{5}\right)\alpha\left(\partial_{i}\beta_{j}-\partial_{k}\beta_{k}\delta_{ij}\right)+\left\{i\leftrightarrow j\right\}.\label{F2}
	\end{eqnarray}
and
	\begin{equation}
	\mathcal{W}^{(3)}=\mathcal{W}_{\alpha\beta^{2}}^{(3)}+\mathcal{W}_{\alpha^{2}h}^{(3)}+\mathcal{W}_{\alpha h^{2}}^{(3)}+\mathcal{W}_{\beta^{2}h}^{(3)}+\mathcal{W}_{h^{3}}^{(3)},\label{W3}
	\end{equation}
with
	\begin{eqnarray}
	\mathcal{W}_{\alpha\beta^{2}}^{(3)} & = & 4\left(c_{1}-c_{5}\right)\alpha\left(2\partial_{i}\beta_{i}\partial_{j}\beta_{j}-\partial_{i}\beta_{j}\partial_{j}\beta_{i}-\partial_{i}\beta_{j}\partial_{i}\beta_{j}\right),\label{W3_abb}\\
	\mathcal{W}_{\alpha^{2}h}^{(3)}&=&-2\left(b_{1}+2\left(c_{1}-c_{5}\right)\right)\alpha^{2}\left(\partial_{i}\partial_{j}h_{ij}-\partial^{2}h_{ii}\right),\label{W3_aah}\\
	\mathcal{W}_{\alpha h^{2}}^{(3)} & = & \alpha\Big(-4c_{5}\partial^{2}h_{jj}h_{kk}+4c_{1}\partial^{2}h_{jk}h_{jk}+4c_{5}\partial_{i}\partial_{j}h_{ij}h_{kk}+4c_{1}\partial_{i}\partial_{j}h_{kk}h_{ij}\nonumber \\
	 &  & \quad-8c_{1}\partial_{i}\partial_{j}h_{ik}h_{jk}-2c_{5}\partial_{i}h_{jj}\partial_{i}h_{kk}+2\left(2c_{1}-c_{5}\right)\partial_{i}h_{jk}\partial_{i}h_{jk}\nonumber \\
	 &  & \quad+4c_{1}\partial_{j}h_{ij}\partial_{i}h_{kk}-2c_{1}\partial_{k}h_{ij}\partial_{i}h_{jk}-2\left(3c_{1}-2c_{5}\right)\partial_{i}h_{ij}\partial_{k}h_{jk}\Big),\label{W3_ahh}
	\end{eqnarray}
	\begin{eqnarray}
	\mathcal{W}_{\beta^{2}h}^{(3)} & = & \left(3c_{1}-2c_{5}\right)\partial_{i}\beta_{i}\partial_{j}\beta_{j}h_{kk}-3c_{1}\partial_{i}\beta_{j}\partial_{j}\beta_{i}h_{kk}+2c_{5}\partial_{i}\beta_{j}\partial_{i}\beta_{j}h_{kk}-2c_{1}\partial_{i}\beta_{j}\partial_{k}\beta_{j}h_{ik}\nonumber \\
	 &  & +4c_{1}\partial_{i}\beta_{j}\partial_{k}\beta_{k}h_{ij}-2c_{1}\partial_{i}\beta_{j}\partial_{i}\beta_{k}h_{jk}-4c_{1}\beta_{i}\partial_{i}\beta_{j}\partial_{j}h_{kk}+4c_{1}\beta_{i}\partial_{i}\beta_{j}\partial_{k}h_{jk}\nonumber \\
	 &  & -2\left(3c_{1}-2c_{5}\right)\beta_{i}\partial_{k}\beta_{j}\partial_{j}h_{ik}+4\left(b_{1}+c_{1}\right)\beta_{i}\partial_{j}\beta_{i}\partial_{j}h_{kk}+2\left(5c_{1}-2c_{5}\right)\beta_{i}\partial_{j}\beta_{j}\partial_{k}h_{ik}\nonumber \\
	 &  & -4\left(b_{1}+c_{1}\right)\beta_{i}\partial_{j}\beta_{i}\partial_{k}h_{jk}-4c_{1}\beta_{i}\partial_{j}\beta_{k}\partial_{j}h_{ik},\label{W3_bbh}
	\end{eqnarray}
	\begin{eqnarray}
	\mathcal{W}_{h^{3}}^{(3)} & = & c_{1}h_{ij}\partial_{i}h_{kl}\partial_{j}h_{kl}-c_{1}h_{ij}\partial_{i}h_{kk}\partial_{j}h_{ll}+4c_{1}h_{ij}\partial_{j}h_{ik}\partial_{k}h_{ll}-2c_{1}h_{ij}\partial_{k}h_{ll}\partial_{k}h_{ij}\nonumber \\
	 &  & +c_{5}h_{ii}\partial_{k}h_{ll}\partial_{k}h_{jj}-\left(5c_{1}-2c_{5}\right)h_{ij}\partial_{k}h_{ik}\partial_{l}h_{jl}-4c_{1}h_{ij}\partial_{j}h_{ik}\partial_{l}h_{kl}\nonumber \\
	 &  & -\left(3c_{1}-2c_{5}\right)h_{ii}\partial_{j}h_{jk}\partial_{l}h_{kl}+2c_{1}h_{ij}\partial_{k}h_{ij}\partial_{l}h_{kl}-2c_{5}h_{ii}\partial_{k}h_{jj}\partial_{l}h_{kl}\nonumber \\
	 &  & +\left(3c_{1}-2c_{5}\right)h_{ij}\partial_{k}h_{jl}\partial_{l}h_{ik}+2c_{1}h_{ij}\partial_{l}h_{jk}\partial_{l}h_{ik}\nonumber \\
	 &  & +3c_{1}h_{ii}\partial_{k}h_{jl}\partial_{l}h_{jk}-c_{5}h_{ii}\partial_{l}h_{jk}\partial_{l}h_{jk}.\label{W3_hhh}
	\end{eqnarray}
	

\bibliography{Gao}

\end{document}